# Laser induced phase transition in epitaxial FeRh layers studied by pump-probe valence band photoemission

Federico Pressacco,[1,2] Vojtěch Uhlíř [3,4], Matteo Gatti[1,5,6], Alessandro Nicolaou[1], Azzedine Bendounan[1], Jon Ander Arregi[3], Sheena K. K. Patel[7], Eric E. Fullerton[7], Damjan Krizmancic[8], Fausto Sirotti[1,9]

[1]Synchrotron SOLEIL, Saint-Aubin, BP 48, F-91192 Gif-sur-Yvette Cedex, France.
[2]Center for Free Electron Laser Science, University of Hamburg, Luruper Chaussee 149, 22761 Hamburg, Germany
[3]CEITEC BUT, Brno University of Technology, Purkyňova 123, 612 00 Brno, Czech Republic
[4]Institute of Physical Engineering, Brno University of Technology, Technická 2, 616 69 Brno, Czech Republic
[5]Laboratoire des Solides Irradiés, École Polytechnique, CNRS, CEA, Université Paris Saclay, F-91128 Palaiseau, France.
[6] European Theoretical Spectroscopy Facility (ETSF)
[7]Center for Memory and Recording Research, University of California, San Diego, La Jolla, California 92093-0401, USA
[8]Instituto Officina dei Materiali (IOM)-CNR Laboratorio TASC, in Area Science Park S.S.14, Km 163.5, I34149 Trieste (Italy)
[9]Physique de la Matiére Condensée, CNRS and École Polytechnique, Université Paris Saclay, F-91128 Palaiseau, France

## Abstract

We use time-resolved X-ray photoelectron spectroscopy to probe the electronic and magnetization dynamics in FeRh films after ultrafast laser excitations. We present experimental and theoretical results which investigate the electronic structure of the FeRh during the first-order phase transition identifying a clear signature of the magnetic phase. We find that a spin polarized feature at the Fermi edge is a fingerprint of the magnetic status of the system that is independent of the long-range ferromagnetic alignment of the magnetic domains. We use this feature to follow the phase transition induced by a laser pulse in a pump-probe experiment and find that the magnetic transition occurs in less than 50 ps, and reaches its maximum in 100 ps.

# Introduction

Intense femtosecond laser excitations are currently the most effective tool to induce rapid modifications in material properties. The absorption of optical energy by the electrons of the material induces a non-equilibrium occupation of the electronic levels. This energy is then delivered to other sub-systems, triggering both structural and magnetic dynamics. A particularly interesting case is when a light pulse brings enough energy to impulsively increase the system temperature above a transition temperature and induce a phase transition (*e.g.* magnetic-nonmagnetic, antiferromagnetic-ferromagnetic, or metal-insulator transitions). Due to the intrinsic characteristic time scales of the interactions between the electrons, spins and lattice, measurements at ultrashort time scales can separate the relative contributions of each sub-system and shed light on which promotes the phase transition.

The application of this approach was first applied to ferromagnetic materials to probe the effects of ultrafast laser excitation on magnetic order, *i.e.* how the electronic excitations are transferred to the spin system[1]. Many theoretical [2][3] and experimental [4][5] efforts have been made to develop a comprehensive model of the laser-induced magnetization dynamics. The first experiments following the disappearance of long-range magnetic order at ultrafast time scales in ferromagnetic Ni were performed using the magneto-optical Kerr effect [6]. In these experiments, the measured quantity is the rotation of the polarization plane of the reflected light, which is proportional to the average magnetization of the sample [7]. These early optical results were confirmed by resonant x-ray based techniques that take advantage of spectroscopic features to analyse the magnetic order in an element specific way and enable measuring the evolution of spin and orbital momenta separately [8]–[10]. X-rays provide numerous techniques to investigate the evolutions and dynamics of long-range magnetic order, such as the resonant x-ray diffraction [1], resonant x-ray reflectivity [11], or single-shot experiments based on coherent scattering using free electron lasers [12][13]. In all cases the signal is either proportional to the net magnetic moment (magnetization) in the measured sample for absorption experiments or related to the variation of the magnetization or charge-magnetic correlations for scattering experiments.

The ultrafast demagnetization process is usually explained as a modification of the electronic structure promoted by the laser pulse absorption. At longer time scales (ps or longer) the electron, spin and lattice systems reach thermal equilibrium approaching or crossing the Curie temperature of the ferromagnetic material followed by remagnetization at longer times scales. However, a complication in interpreting magnetization data is the misalignment of magnetic domains in an external field. This is particularly true in the remagnetization process where alignment of domains can occur at precessional time scales (typically > 100 ps). Thus a preferable approach is following the response of corresponding signatures in the electronic structure instead of the spatially averaged magnetic order. However, only a few experiments have been performed to measure directly the electronic distribution in exchange-split valence states [14] or spin state of the electrons excited in the empty bands [15], [16]. The reason can be identified in the space charge affecting the high resolution ARPES experiments [17] and in the challenges of performing time and spin resolved photoemission spectroscopy [18].

In this paper, we present the results of X-ray Photoemission Spectroscopy (XPS) experiments performed on FeRh thin films that tracks the ferromagnetic signature in the valence band structure across the magnetic phase transition. FeRh present a particularly interesting system for exploring the interplay of structural, magnetic and electronic phase transitions in metallic systems [19]–[22]. Ordered FeRh alloys of about 50 at % Rh undergo a first-order phase transition from a low-temperature antiferromagnetic (AFM) to a high-temperature ferromagnetic (FM) phase upon heating above 360 K [19], [23] . In the low temperature AFM phase, FeRh is a G-type antiferromagnet where Fe atoms carry a moment of ±3.3 $\mu_B$, while Rh atoms possess a negligible magnetic moment. In the FM phase, a collinear spin configuration is found, with local ferromagnetically coupled moments of 3.2 $\mu_B$ and 0.9 $\mu_B$ for Fe and Rh sites, respectively [24]–[27]. The two phases are isostructural but show sizable differences in the cell size, inducing a volume expansion of about 1 % [28], and in the electronic structure [29]–[31]. It has been recently shown that the phase transition in FeRh can be induced by laser excitation although there is an ongoing debate on the evolution of the FM phase after laser heating[32]–[35].

Using XPS, we have followed the AFM-FM transition by measuring the increase in the electronic spectral weight at the Fermi level induced by laser pulses over 4.5 ns with a temporal resolution of 50 ps. Comparing our experimental results with *ab initio* calculations of the electronic structure applied to the AFM and FM phases, we could correlate this spectral feature at the Fermi level with the different magnetic order of the two phases as observed by Lee and co-workers [31]. We used this feature to follow the phase transition both in static and dynamic regimes. A parameter-free theoretical description is an essential ingredient to be confident in the attribution of the electronic structure features. Moreover, a detailed description of the electronic structure is the natural starting point for future developments aiming to describe electronic excitations induced by short laser pulses and the relaxation processes.

## METHODS

## SAMPLES

Epitaxial FeRh layers of 50-nm thickness were prepared on MgO(100) substrates by dc magnetron sputtering using an equiatomic target. The films were grown at 725 K and post-annealed at 1070 K for 45 minutes. The films are subsequently cooled down in the deposition chamber and protected by a 2-nm-thick Pt capping layer. For XPS measurements the capping layer was removed by few cycles of light Ar sputtering and annealing [36] . We performed Vibrating Sample Magnetometry (VSM) measurements of the net magnetic moment to identify the phase transition temperatures[1] upon heating ($T_{FM}$) and cooling ($T_{AFM}$), see Fig1(a). The measurement is performed in an in-plane applied magnetic field of 1 T inducing a shift of the thermal hysteresis loop of about −8 K [37] [38]. The width of the hysteresis is 12 K, which indicates a homogeneous sample with low variation in stoichiometry.

---

[1] Defined as the temperature at which the magnetization reaches half of the maximum value.

It is worth noting that the magnetization is not zero at room temperature where the system is expected to be in the AFM phase. However, it has been demonstrated that a top FM layer in the AFM FeRh film is induced in case of a Rh-terminated free surface [36], at an interface with a capping layer [39], or with the MgO substrate [40].

**Experimental setup**

The XPS experiments were performed using the UHV-Photoemission experimental station of the TEMPO beamline at the SOLEIL synchrotron radiation facility. The measurement chamber is equipped with a 2D Scienta SES 2002 analyser with an angular resolution better than 0.2 degrees in the angular range of 15 degrees in the horizontal plane. The detection system is equipped with a delay line detector, which allows performing pump/probe time resolved experiments [41], [42]. The photon energy ranges from 50 to 1500 eV, with an energy resolution $E/\Delta E$ better than $10^4$. For this experiment we used the photon energy of 120 eV. An electromagnet, installed in the back of the sample, allows inducing an in-plane magnetization in the horizontal plane. The sample magnetization can be selected before photoelectron spectroscopy experiments, which are performed in remanence at zero applied magnetic field.

We use 50-fs laser pulses with a central wavelength of 800 nm for the optical excitation, generated in the TEMPO optical hutch by a Coherent REGA 9050 and subsequently focused at the sample position to a spot size of about 200 microns full-width at half maximum (FWHM), in order to be superposed on the soft x–ray probe spot. Because of the 25-degree incidence angle with respect to the sample normal, the circular focal spot results in a projected elliptical shape of 200 microns FWHM and 550 microns in the vertical and horizontal planes, respectively. The maximum power corresponds to a fluence of 5 mJ/cm$^2$, which is sufficient to induce the AFM-FM phase transition in FeRh thin films [32].

The laser pulses are synchronized and delayed with respect to the isolated bunch in the SOLEIL time structure. The latter pulse generates the photoelectrons, and has a temporal width of 50 ps, which determines the temporal resolution of our experiment. The laser repetition rate was 141 kHz, corresponding to six periods of the synchrotron revolution, *i.e.* the measuring synchrotron pulse is followed by five isolated synchrotron pulses separated by about 1.2 µs before a new excitation. Because of the short (nanosecond) relaxation time observed for the laser-induced generation of FM order in FeRh, photoelectron intensity excited by the five unexcited synchrotron pulses can be used to normalize the photoemission intensity at different delays between laser and synchrotron. The energy supplied by the laser together with the thermal insulation of the sample holder and the UHV environment, determines the static heating of the whole system. To investigate the phase transition dynamics, the system is in the AFM phase before the laser excitation. In order to prevent the drift of temperature above the transition temperature due to static heating, we kept the sample holder at a fixed temperature $T_0$=220 K during experiments.

Time-resolved magneto-optical Kerr effect (TR-MOKE) experiments were performed in the laser hutch of the TEMPO beamline on the sample protected by a thin Pt layer in air.

During the experiments a magnetic field of 0.1 T was applied to the sample. The probe pulse consisted in a laser pulse with a central wavelength of 400 nm (second harmonic of the fundamental wavelength). In this case, the measurement is less surface sensitive due to the larger penetration depth of the probe beam (approximately 10 nm).

**Theoretical tools**

Theoretical Kohn-Sham band structure was obtained with density functional theory in the local density approximation[43]. We have adopted the experimental lattice parameters for both phases [28]. We have used Hartwigsen-Goedecker-Hutter norm-conserving pseudopotentials [44] in a plane-wave basis approach [45], where Fe 3s 3p and Rh 4s 4p semicore states have been explicitly treated as valence electrons. The calculations have been converged with a 115 Hartree cut-off energy and 20x20x20 Monkhorst-Pack k-point grid, and the results are reported in Figure 1b).

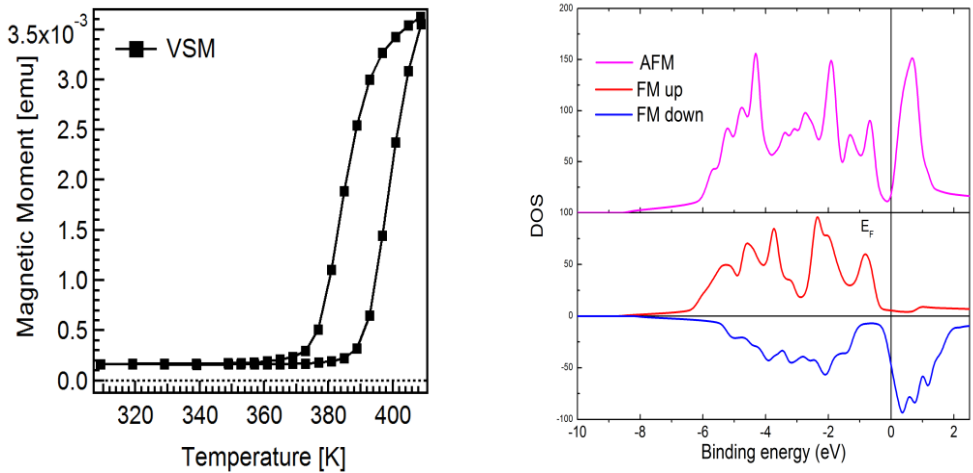

Figure 1: *(a) The magnetization of the 50 nm thick FeRh film measured by VSM as a function of temperature. We compensated for the shift induced by the applied magnetic field of 1 T during the measurement by translating the curve 8 K toward higher temperatures. The non-zero signal at room temperature is due to the presence of a ferromagnetic layer at the interfaces with the substrate and capping layer. From the hysteresis we extracted two transition temperatures of 387 K ($T_{FM}$) and 375 K ($T_{AFM}$). (b) Calculated spin-polarized density of states (DOS) for the two phases. The zero of the energy axis is set at the Fermi level. In the AFM phase spin up (majority) and spin down (minority) DOS are the same.*

**RESULTS and DISCUSSION**

**Electronic structure to identify FM phase**

In order to identify the fingerprint of the FeRh magnetic state in the electronic structure, we investigated the band structure of the two phases both theoretically and experimentally. One can better understand the differences in the electronic structures between the two phases by considering the momentum integrated spectrum [46], [47], and focusing the attention in the region close to the Fermi edge. The theoretical and experimental spectra measured in the energy range of 2 eV across the Fermi level are compared in Fig. 2. Figure 2a) shows the calculated total DOS for the two phases, while the XPS signals are presented in 2b). In agreement with previous results from literature [26], [29], [30], [48]–[50], the calculations show a clear difference in the intensity of the valence band near the Fermi level. The modification of the photoemission intensity in the XPS signal in the binding energy range 0.5 eV to 1 eV is smaller than the one in the total DOS because the latter accounts also for electrons emitted outside the acceptance angle of the spectrometer. In any case, the change of the spectral weight near Fermi energy is clearly visible also in the measured photoemission spectra. The theoretical simulations furthermore allow us to identify the origin of this spectral change between the two phases. By comparing the spin-resolved DOS, see Fig. 1(b), we notice that the peak at the Fermi level in the FM phase is strongly spin polarized and is actually originating from the tail of a structure in the unoccupied states (hence not accessible in photoemission) that is present only in the spin-minority DOS. In the AFM phase, instead, a dip in the DOS opens at the Fermi level. We can therefore conclude that the peak at the Fermi level is a clear fingerprint of the FM state of the system.

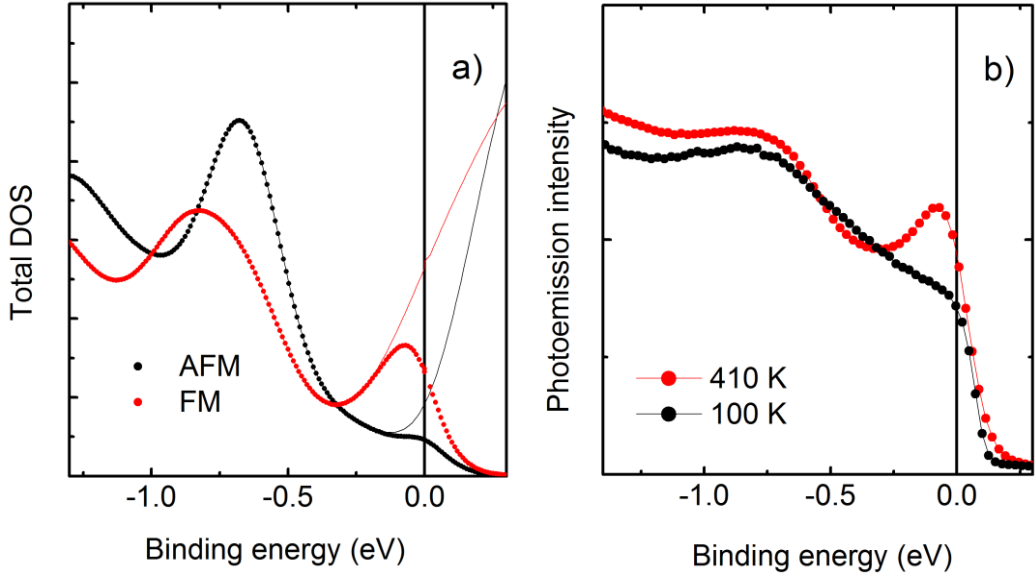

**Figure 2**: *a) Density of states calculated near the Fermi level for the FM and AFM phases are indicated by red and black solid lines, respectively. The symbols are obtained by multiplication by a Fermi function corresponding to temperatures of panel b). b) Experimental lineshapes obtained by integrating the ARPES map across 10 degrees around Γ*

To follow the evolution of this peak across the phase transition, we adopt a fit procedure to extract the most relevant parameters from the experimental spectra, namely the intensity of the FM peak and the temperature T. In Fig. 3 (a) we show the fit results for the spectrum measured at 430 K. We have reproduced the spectral shape using a function built as the product of a Fermi function and the sum of a Gaussian shape representing the FM peak and a Doniach-Sunjic function (B1) centred at about 0.7 eV binding energy. The Fermi function (grey area in Fig. 3(a)) identifies the electronic temperature of the system. The Doniach-Sunjic function takes into account the intensity contribution from several deeper bands and the asymmetry of the peak due to secondary electron contributions. This is a pure phenomenological description, which leads to a good convergence of the fit procedure and allows us to quantify the intensity modifications taking place in the binding energy region near the Fermi level. In order to reduce the number of free parameters, we kept constant the width of the FM peak together with the width and asymmetry of the B1 band.

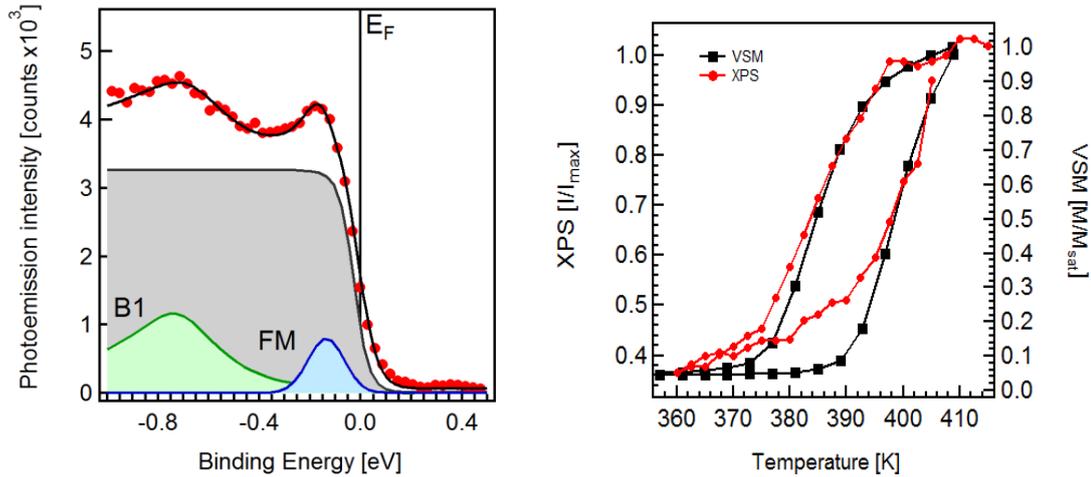

*Figure 3: (a) Fitting procedure applied to the spectrum measured at 430 K when the system is in the FM phase. The measured photoemission intensity (red dots) is reproduced by adding B1 and FM components multiplied by a Fermi function. (b)The normalized intensity of the FM peak extracted from the fit is presented as a function of the temperature (red curve) and compared with the relative magnetic moment measured in VSM (black curve).*

We performed a temperature cycle from 430 K down to room temperature and back to 400 K and recorded a spectrum every 2 K. Using the fit procedure described above, we could effectively estimate the intensity of the peak close to the Fermi edge and its dependence on temperature. The result of the fit is reported in Fig 3(b) (red symbols) together with the VSM data (black symbols) used for comparison. The XPS experiment is performed without applied magnetic field and the temperature scale of the VSM was corrected by 8 K [37]. One can clearly observe the same hysteretic behaviour with the same transition temperatures extracted from the two experiments. The growth of the FM peak intensity beginning at lower temperatures can be attributed to the formation of small FM domains which contribute to the XPS signal. Due to their small dimensions, exchange effects with the AFM matrix freeze the magnetization in a random orientation, which will then average to zero in the VSM experiment [51]. On the other side, their electronic properties will be that of the FM phase and hence detected in the XPS signal. The comparison between XPS and VSM data confirms that the density of states in the valence band close to the Fermi edge can be used to quantitatively follow the AFM-FM phase transition of FeRh. Thus we see that the pathway to measure the fraction of FM order in FeRh via XPS possesses a relevant distinction from those measurement strategies relying on the detection of magnetization. While the valence band feature reported here is only sensitive to the existence of FM order, it is at the same time independent of the specific magnetization orientation in the entire probed region, thus not requiring magnetization alignment of FM domains by means of an applied magnetic field. Therefore in the following we will use this spectral feature also to investigate the AFM-FM phase transition induced by the ultrafast laser excitation.

**Dynamics of the AFM / FM phase transition by probing the valence band**

Former investigations addressed the dynamics of the phase transition in FeRh by measuring the anomalous change in the lattice expansion [35] with x-ray diffraction, or comparing it to the magneto-optical response [34]. The work of Mariager *et al.* [34] speculated that the magnetic response measured with the magneto-optical Kerr effect, given by the alignment of the magnetic moments to the external field, is slower than the structural response since the latter is independent of the magnetization orientation. In a similar way, measuring by XPS the change of the electronic structure associated with the different magnetic phases does not require to align the magnetic moments to an external field. The photoelectron spectroscopy experiment performed in the present geometry with linearly polarized light is not sensitive to the magnetization orientation.

We performed the XPS measurement for delays ranging from -600 ps to 4 ns. In Fig. 4a, we present the color map of the photoemission intensity measured in the near region of the Fermi level as a function of the delay between the laser pump pulse and the synchrotron probe. In Fig. 4b, we show selected spectra (solid circles) and fit results (solid lines) before and after the laser excitation. In fitting the time resolved spectra, as we did for the temperature dependent experiment, we keep the energy difference between both bands and the Fermi edge fixed but we allow rigid shifts of the binding energy in order to compensate for the shift induced by the space charge [17].

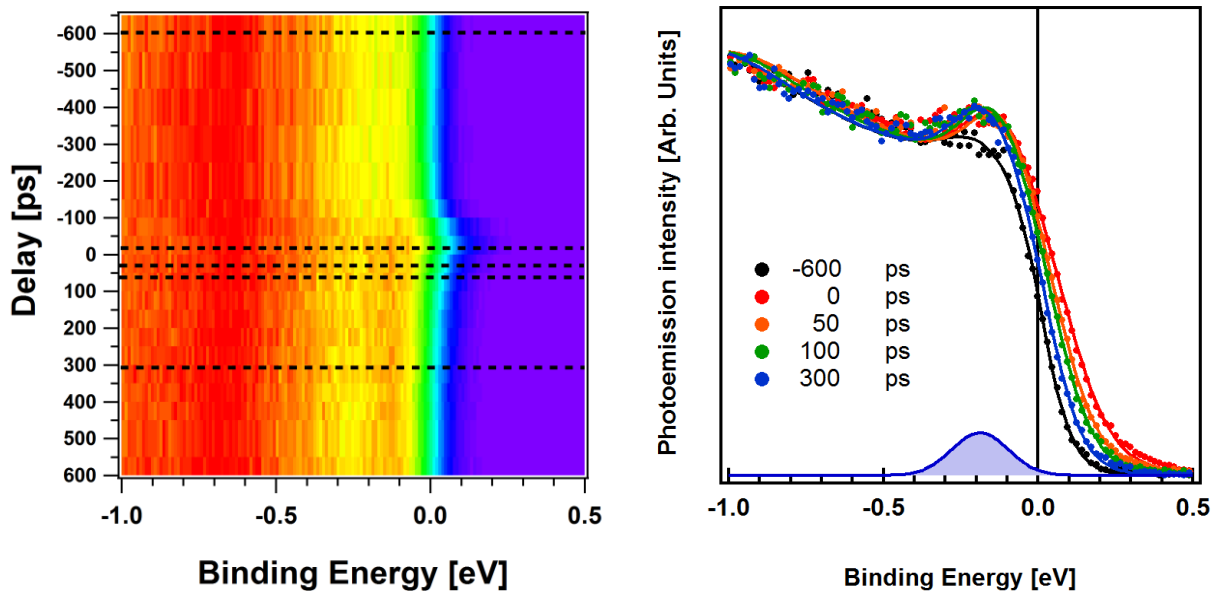

*Figure 4: a) Color map representing the measured photoemission intensity as a function of the binding energy (horizontal scale) and the delay between the synchrotron and the laser pulses. The*

*dashed lines indicate the spectra presented in **b**) Selected spectra measured in the binding energy region near the Fermi level with 100-eV photon energy (solid dots). The delay between the laser and synchrotron radiation pulses is indicated in the figure. The Gaussian at the bottom represents the intensity associated with the FM phase appearing near the Fermi level measured at 300 ps.*

The unperturbed spectrum (black symbols) is measured 600 ps before the arrival of the laser pulse at time t=0. The FM peak intensity corresponds to the one measured in static conditions at room temperature. This confirms that the cooling of the sample effectively prevented the temperature drift due to static laser heating. In the spectra of Fig. 4b, the position of the Fermi edge of the spectrum at t=0 is shifted toward positive binding energies because of the space charge created by the pumping femtosecond laser pulse [17], [52]. Space charge due to the intense infrared laser pump is expected to disturb the photoelectron spectroscopy experiment, but for reduced power densities, spectral deformations can be controlled and high-quality ARPES data can be obtained [14], [15], [41].

All spectra measured at positive time delays show an increase in the electronic density close to the Fermi edge, which is characteristic of the FM phase. The Gaussian profile depicted in Fig. 4 marks the position of the peak associated with the FM phase (in this case retrieved from the spectrum measured at 300 ps). The slope modification of the Fermi edge is also clearly visible at positive times and can be associated with the temperature increase. The parameters describing the transition to the FM phase and the relaxation to the AFM one are presented in Fig. 5. The sample holder was kept at 220 K while the sample temperature can be extracted from the fit procedure described previously. It is plotted against time in the top panel. Before *t=0*, the electronic temperature is about 350 K. We can measure the static heating induced by the laser pulse associated with the heat transfer in the sample environment: it is about 110 K. In metallic systems, the energy transfer between electrons and phonons is completed within the first few ps [53]. Since our experiment has a time resolution of 50 ps, we can assume that the extracted value of the electronic temperature corresponds to the lattice temperature of the system. The only exception is *t=0*, where space charge effects can affect the measured value of 650 K for an unknown time interval within the 50 ps time window accessible from the synchrotron pulse width.

The following temperature decay is fit by two exponentials as shown in Fig. 5:

$$f(t) = y_0 + A_1 e^{-\frac{t-t_0}{\tau_1}} + A_2 e^{-\frac{t-t_0}{\tau_2}}$$

where $\tau_i$ are the two time constants, $t_0$ the temporal overlap between pump and probe, and $A_i$ are the relative intensities of the exponentials. From the fit we retrieve two different time constants $\tau_1=133\pm24$ ps and $\tau_2=1.2\pm0.9$ ns. They are in good agreement with the values associated with the thermal diffusion within the film and the substrate, respectively [54]. It is worth noting that the temperature drops below $T_{AFM}$ after 1700 ps.

In the bottom panel of Figure 5, we present (blue circles) the amplitude of the FM spectral intensity (left scale) as a function of the delay between the laser and synchrotron pulses. The transition is immediately measured at t=0, where the intensity of the FM peak increases by more than a factor of 3. Due to our limited temporal resolution (black curve in Fig. 5), we can only demonstrate that the electronic structure reacts to the transition faster than 50 ps, and reaches its maximum in 100 ps. It is faster than the time resolved measurement of the Kerr rotation (TR-MOKE) performed on the same sample and with the same pump energy in an all-optical pump-probe experiment indicated by the green line (right scale). As previously reported [32]–[34], the Kerr rotation shows a slow increase and the signal reaches its maximum after 500 ps indicative of domains in FM phase slowly aligning after the phase transition. .

The short time scale of XPS is consistent with or faster than the domain growth of the ferromagnetic phase observed with XRD technique [34], [35]. Because of the shorter probing depth of photoelectron spectroscopy and the reasonable assumption that the nucleation of the FM phase starts at the free surface, time-resolved XPS is well suited to study fundamental limits of laser-induced phase transitions [5] and to describe the energy transfer from electron excitations to spin and lattice systems.

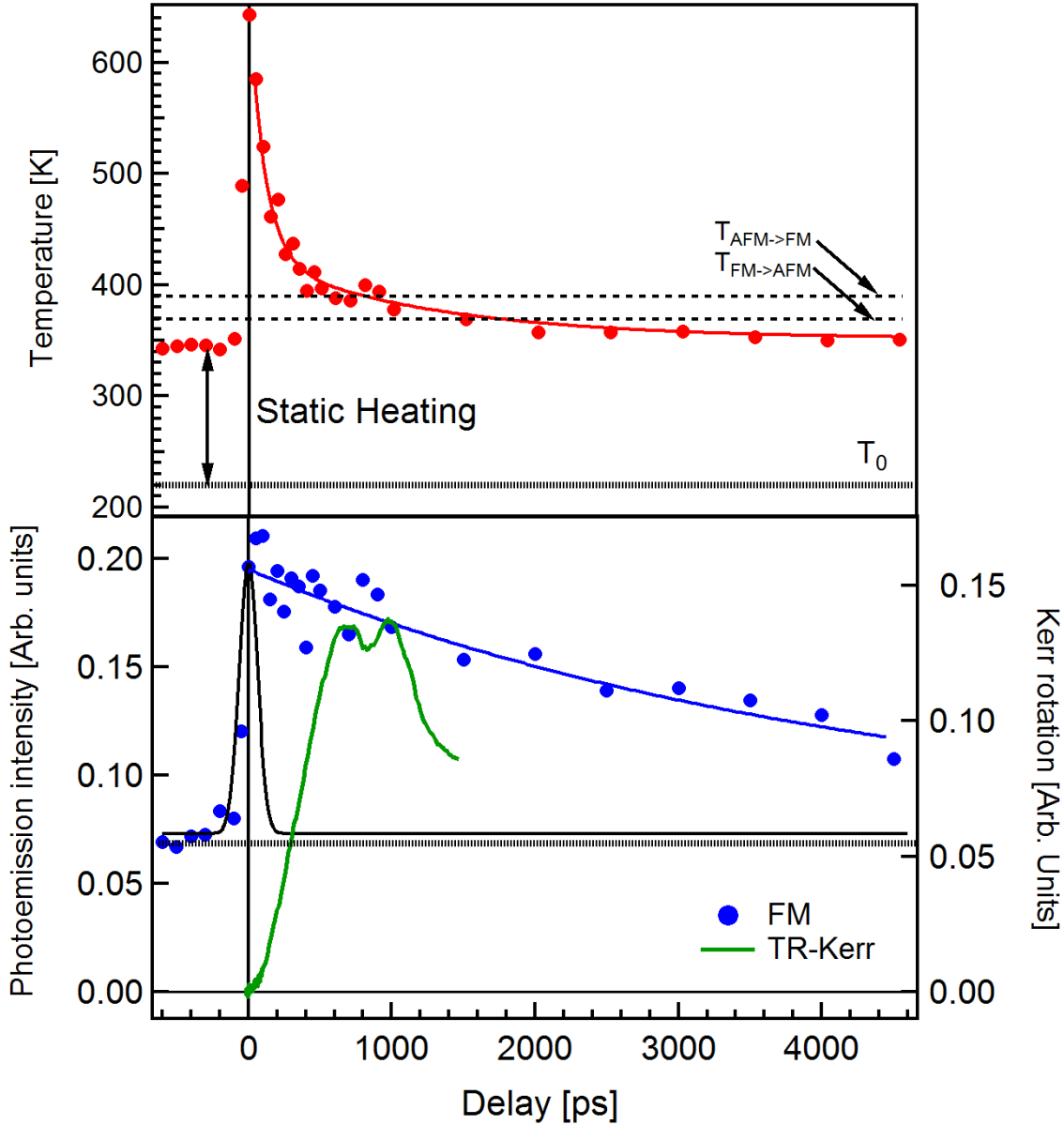

*Figure 5: Top)* Time dependence of the sample temperature (left scale) extracted from the coefficient of the Fermi function in the fitting procedure. $T_0$ is the sample holder temperature. The stating heating due to the high-frequency laser pulses and the critical temperatures for the phase transition are indicated. The solid red curve is the result of the fit as described in the text. *Bottom)* Amplitude of the FM band signal (blue filled circles) upon the fs laser excitation. The black line shows the x-ray pulse time profile, while the blue solid line is the result of the fit. The green line (right scale) is the TR-MOKE signal measured in an applied magnetic field of 200 mT.

The FM peak intensity in XPS starts to decrease after 200 ps and the relaxation is not completed even after 4.5 ns. The TR-MOKE signal relaxation to the AFM phase starts after about 1 ns and it is completed within 2 ns. It is interesting to notice that even if after 1.7 ns the system temperature goes below the transition temperature to the AFM phase under static conditions, the FM signature is still clearly visible. This relaxation time is slower than that observed using other all-optical pump probe experiments [32] and even time-resolved XMCD studies [55]. This long tail can be attributed to the presence of small randomly-oriented FM domains in the AFM phase as observed in the temperature hysteresis discussed previously. They could remain longer in the system and keep a FM phase contribution in the photoemission signal. This would not be present even in the XRD experiment because too small domains would give broad diffraction peaks.

## Conclusions

In conclusion, we have identified a sharp peak in the density of states of FeRh FM phase close to the Fermi level which can be used to discriminate between the AFM and FM phases. Theoretical Kohn-Sham electronic structure obtained with density functional theory in the local density approximation is in a good agreement with photoemission experiments. The photoemission intensity in the vicinity of the Fermi level measured as a function of temperature through the AFM to FM phase transition is in excellent agreement with the thermal magnetic hysteresis observed by VSM experiments on the same layer. A difference observed in the rise of the hysteresis curves can be attributed to small pinned FM domains present close to the surface at lower temperatures, which are not visible in the magnetization dependent experiment.

We have measured the appearance of the FM peak with time resolved X-Ray Photoemission spectroscopy (XPS) after a 50 fs laser excitation in FeRh epitaxial layers. A fitting procedure applied to the high resolution spectroscopy signals reproduced the increasing of the characteristic peak intensity associated with the FM phase as well as the slope of the DOS at the Fermi level associated with the sample temperature. In pump/probe photoelectron spectroscopy experiments the transition to the FM phase is fully completed within 100 ps since the laser excitation, in agreement with published XRD results. The reestablishment of the AFM phase is considerably slower, on the time scale of few ns. We attribute this difference to the presence of small FM domains which would contribute to the electronic structure experiments, but not to those measuring magnetization differences. This is confirmed by the comparison with all-optical time-resolved MOKE experiments performed under similar excitation conditions. The observed modifications in the electronic structure do not need a magnetization alignment of the sample and for this reason it can be used to investigate the ultimate time scale of the AFM to FM phase transition in FeRh.

## Acknowledgements

This work has been supported by the excellence cluster 'The Hamburg Centre for Ultrafast Imaging - Structure, Dynamics and Control of Matter at the Atomic Scale' of the Deutsche Forschungsgemeinschaft and has received funding from the EU-H2020 research and innovation programme under grant agreement No 654360 having benefitted from the access provided by Synchrotron SOLEIL in Saint-Aubin - BP 48 91192 GIF-sur-YVETTE (France), within the framework of the NFFA-Europe Transnational Access Activity". This research was supported by a Marie Curie FP7 Integration Grant. Computation time was granted by GENCI (Project No. 544). Research at CEITEC BUT is funded by the Grant Agency of the Czech Republic (grant no. 16-23940Y) and the Ministry of Education, Youth and Sports of the Czech Republic under the projects CEITEC 2020 (LQ1601) and CEITEC Nano (LM2015041). Research at UCSD on sample growth and materials characterization supported by U.S. Department of Energy (DOE), Office of Basic Energy Sciences (Award No. DE-SC0003678).